\documentclass[showpacs,twocolumn,a4paper,10pt,prl]{revtex4-1}
\usepackage[utf8x]{inputenc}
\usepackage{amsmath}
\usepackage{amsfonts}
\usepackage{amssymb}
\usepackage{graphicx}

\begin{document}

	\title{Disentangling sources of anomalous diffusion}

	\author{Felix Thiel}\affiliation{Institute of Physics,
	  Humboldt University Berlin, Newtonstr. 15, 12489 Berlin, Germany}
	\author{Franziska Flegel}\affiliation{Institute of Physics,
	  Humboldt University Berlin, Newtonstr. 15, 12489 Berlin, Germany}
	\author{Igor M. Sokolov}\affiliation{Institute of Physics, Humboldt
	  University Berlin, Newtonstr. 15, 12489 Berlin, Germany}
	\date{\today}

	\begin{abstract}
		We show that some important properties of subdiffusion of unknown origin (including ones of mixed origins) can be easily assessed when finding the ``fundamental moment'' of the corresponding random 
		process, i.e., the one which is additive in time.
		In subordinated processes, the index of the fundamental moment is inherited from the parent process and its time-dependence from the leading one. 
		In models of particle's motion in disordered potentials, the index is governed by the structural part of the disorder while the time dependence is given by its energetic part. 
	\end{abstract}
	\maketitle

		Anomalous diffusion has become a much discussed topic in the recent years.
		It is defined as the random motion of some object, in which the mean squared displacement (MSD) grows not linearly in time, 
			but follows a different power law, 
			i.e. $\left\langle X^2(t) \right\rangle \sim t^{\alpha} $, with $\alpha \ne 1$.
		Here we concentrate on subdiffusion, $\alpha < 1$.
		Subdiffusion is a behaviour emerging in many situations with most recent examples being pertinent to transport inside biological cells 
			\cite{Bronstein2009,Jeon2011,Weber2010,Golding2006,Seisenberger2001} or on their membrane \cite{Kusumi2005,Weigel2011}, see  \cite{Hoefling2013} for a review.
		Physically, anomalous diffusion may have very different causes \cite{Haus1987,Bouchaud1990}. 
		It may appear due to the internal degrees of freedom in viscoelastic systems, 
			due to labyrinthine surroundings or due to traps in disordered systems, etc., each case 
			needing a different mathematical instrument for its description \cite{Sokolov2012}. 
		In an experiment, it is not always easy to decide, from the data or from a priori considerations, 
			what situation applies, and what mathematical instrument has to be used.
		Very often, the results of the measurement or of computer simulations giving some properties of a random process $X(t)$, 
			are simply fitted to an ad hoc theoretical model, in the best case some statistical tests are used to check 
			whether the results comply with the prediction of one of the models chosen from a relatively small repertoir of alternatives.
		Here it is necessary to note that subdiffusion may have mixed origins \cite{Meroz2010}, and that such processes were indeed observed in experiments \cite{Weigel2011}, 
			which makes finding the corresponding model even more complicated.
		To understand the situation at hand it is therefore necessary to be able to ``deconstruct'' it, 
			i.e. to classify the process without comparing it with a preexisting theoretical model. 

		In what follows we concentrate on random processes which can be either considered as the ones subordinated 
			to processes with stationary or uncorrelated increments in systems without disorder, 
		or can be approximated by those after averaging over realizations of disorder in disordered systems. 
		For the processes which can be described in the language of subordination, 
			the procedure corresponds to separation of the contributions of the parent and the leading process, 
			and gives us the possibility to obtain some important properties immediately from experimental or numerical data. 
		For the case of diffusion in disordered potential landscapes (which does not in general reduce to subordinated processes in low dimensions) 
			one can clearly separate the structural and the energetic component of the disorder. 

		We first present the main quantity which we consider suitable for this task: 
			the fundamental moment of the process, and elucidate the corresponding notion for the case of subordination of a process 
			with stationary increments under the operational time given by a general process with non-negative increments. 
		Then we turn to disordered systems and give a short summary of two relevant physical situations: of the energetic and of the structural disorder, 
			and show how the method proposed allows for distinguishing the contributions of the corresponding disorder types. 
		The procedure discussed here works on the level of ensemble means; the single-trajectory versions have to follow slightly different lines, 
			and will not be considered here. 
		
	\textbf{i) Processes subordinated to a process with stationary increments.} 
		The subordination models, like the famous continuous time random walks (CTRW), 
			consitute the class for which our procedure is the simplest to apply. 
		Let us thus consider a process subordinated to a random process $X(u)$ (parent process) \textit{with stationary increments} 
			under some operational time $ u = \tau(t)$, i.e. the random process 
		\[
			 X(t) = X(\tau(t))
		\]
		starting at $x=0$ for $u=0$, where the increment process $\Delta X(u_1,u_2) = X(u_2)-X(u_1)$ is a stationary random process in $u$.
		Due to stationarity we are allowed to put $\Delta X(u_1,u_2) = \Delta X(u_2 - u_1)$. 
		The process $\tau(t)$ (leading process) maps the physical time $t$ to the operational time $u$. 
		The process $\tau(t)$ is a general random process with non-negative increments (which guarantees the causality of the model). 
		Let $P(x,u)$ be the PDF of the increment process.
		For simplicity we take $P(x,u)$ to be symmetric: $P(x,u) = P(-x,u)$. 
		Now let us assume that this PDF scales as a function of the operational time lag $u$,
		\begin{equation*}
				P\left( x, u \right) 
			= 
				\frac{1}{u^{\alpha/2}} 
				f\left( \frac{x^2}{ u^{\alpha} } \right).
		\end{equation*}
		No other restrictions are assumed. 

		Let us concentrate of the (generalized) absolute moments of $\Delta X(u)$:
		\[
				\langle |\Delta X(u)|^\gamma \rangle 
			=
				\int_{-\infty}^\infty |x|^\gamma 
				P(x,u) \mathrm{d}x 
			= 
				A_\gamma u^{(\gamma \alpha)/2},
		\]
		with a prefactor $A_\gamma$ depending on the exact form of the PDF of the increment process. 
		We note that for $\gamma = \gamma_F = 2/\alpha$ this moment (averaged over realizations of the parent process $X(u)$ for given $u$) is linear in $u$ and therefore additive in $u$. 
		Thus, taking $\gamma = 2 / \alpha$ and any three operational time instants $0 = u_0 \le u_1 \le u_2$, we have
		\[
				\langle |\Delta X(u_1,u_2)|^{\gamma_F}  \rangle 
			=
				A_{\gamma_F} (u_2-u_1)
			\text{,}
		\]
		so that
		\[
				\langle |\Delta X(u_0,u_2)|^{\gamma_F}  \rangle 
			= 
				\langle |\Delta X(u_0,u_1)|^{\gamma_F}  \rangle +
				\langle |\Delta X(u_1,u_2)|^{\gamma_F} \rangle
			\text{.}
		\]
		We now average this equation over the distribution of the operational times pertinent to three instants of clock time $0 = t_0 \le t_1 \le t_2$: $u_1=\tau(t_1)$ and $u_2=\tau(t_2)$.
		Passing from operational to clock times, we obtain
		\[
				\langle |\Delta X(\tau(t_1),\tau(t_2))|^{\gamma_F}  \rangle_{\tau_1,\tau_2} 
			=
				A_{\gamma_F} \left[ \langle \tau(t_2) - \tau(t_1) \rangle_{\tau} \right],
		\]
		which -- due to additivity -- translates to 
		\begin{equation}
				\langle |\Delta X(t_0,t_2)|^{\gamma_F}  \rangle 
			=  
				\langle |\Delta X(t_0,t_1)|^{\gamma_F}  \rangle 
				+ \langle |\Delta X(t_1,t_2)|^{\gamma_F}  \rangle 
		 \label{fundamental}
		\end{equation}
		(where now a double average over the realizations a parent and of a leading process is performed). 
		Thus, the moment of index $\gamma_F = 2/\alpha$ which was additive in operational time stays additive in the clock time. 
		We stress that $\langle |\Delta X(t,t')|^{\gamma_F} \rangle$ does not have to grow linearly in $t'- t$,
			and it's dependence on it's arguments might be quite complex.
		For example, for simple Brownian motion, or for a simple random walk as a parent process, this is the second moment. 

		We will call the  index $\gamma_F$ the fundamental exponent of our random process, and the moment of index $\gamma_F$ its \textit{fundamental moment}. 
		The index of the fundamental moment of the subordinated process is inherited from the parent process. 
		Having enough realizations of the precess (in all our simulations 2048 trajectories were used) the index $\gamma_F$ can be found by solving Eq. (\ref{fundamental}) numerically. 
		When this index is known, we get
		\begin{equation}
				\langle |X(t)|^{\gamma_F}  \rangle 
			= 
				A_{\gamma_F} \langle \tau(t) \rangle
			\label{TimeDep}
		\end{equation}
		and restore the time dependence of the first moment of the subordinator provided $A_{\gamma_F}$ is known. 
		If, additionally, we have the knowledge of all $A_\gamma$ we can see that 
			$\langle |X(t)|^{2\gamma_F}  \rangle = A_{2\gamma_F} \langle \tau^2(t) \rangle$ and so on, 
			which allows us to restore the sequence of the moments of the subordinator.
		If, on the contrary, the moments of subordinator are known as functions of time, 
			the sequence of generalized moments of the parent process can be obtained.  

	\textbf{ii) Parent process with uncorrelated increments.} 
		In the cases typically considered as examples of subordination, like CTRW, the leading process is taken to be independent from the parent one. 
		The situations in which $\tau(t)$ does depend on $X(u)$ (e.g. through its increments) will be termed ``quasi-subordination''.
		Some of them can be considered on the equal footing. 
		CTRW is a process where the increments of the parent process are both stationary and uncorrelated \cite{Sokolov2012}. 
		This second property by itself is enough for applicability of our method, 
			even if the increments were non-stationary and dependend on $X$.
		Such a situation is realized e.g. in random trap models \cite{Haus1987}. 
		For models with uncorrelated increments the fundamental moment is always the second one.
		This is easily seen from
		\begin{eqnarray*}
				\langle |\Delta X(u_0,u_2)|^2 \rangle 
			&=&
				\langle |\Delta X(u_1,u_2)|^2 \rangle 
				+ \langle |\Delta X(u_0,u_1)|^2 \rangle  \\
			& &
				+ 2 \langle [X(u_2) - X(u_1)] [X(u_1) - X(u_0)] \rangle.
		\end{eqnarray*}
		If the increments are uncorrelated, the last term vanishes and Eq. (\ref{fundamental}) with $\gamma_F = 2$ is recovered.
		By the same arguments as above, the additivity still holds after substitution of the operational times with the clock ones. 
		
\begin{figure}
			\includegraphics[width=0.4\textwidth]{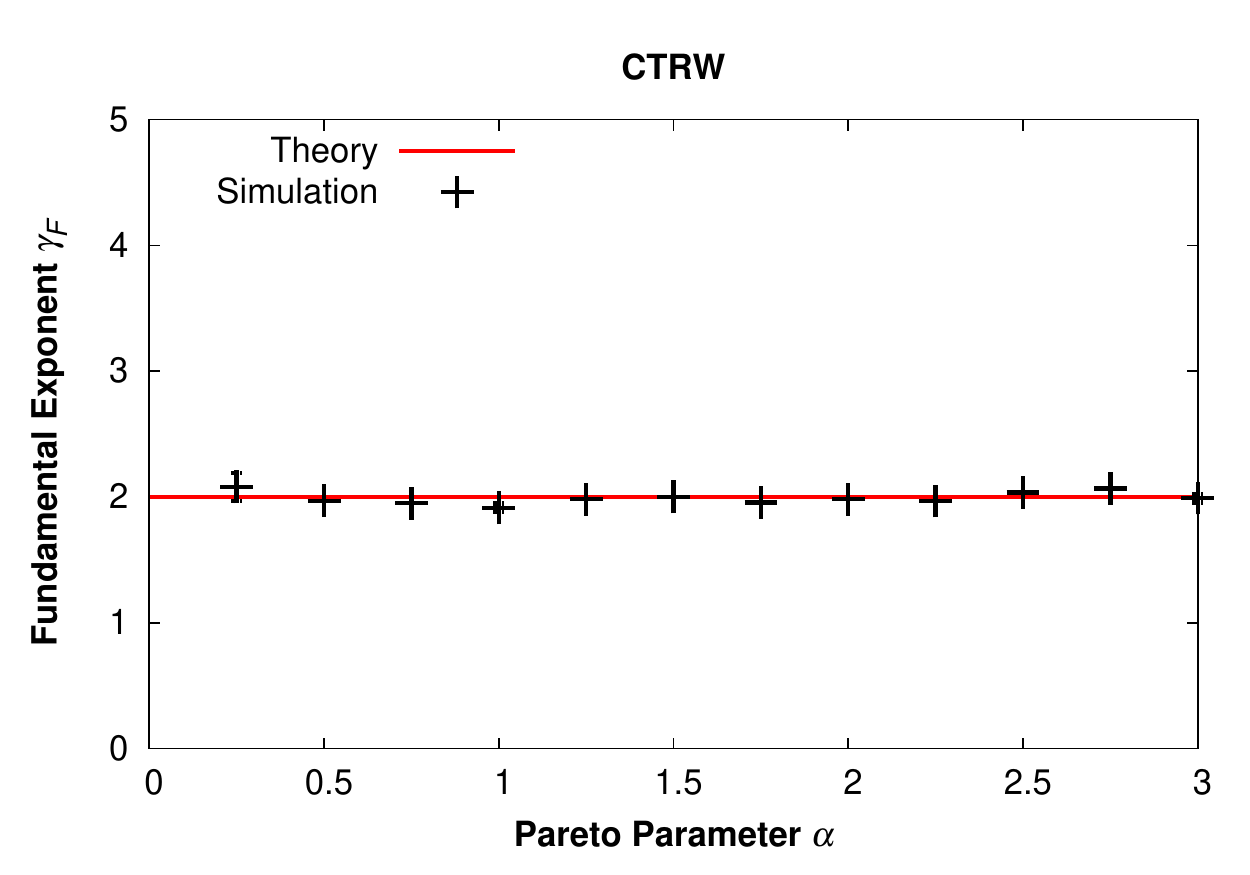}
			\includegraphics[width=0.4\textwidth]{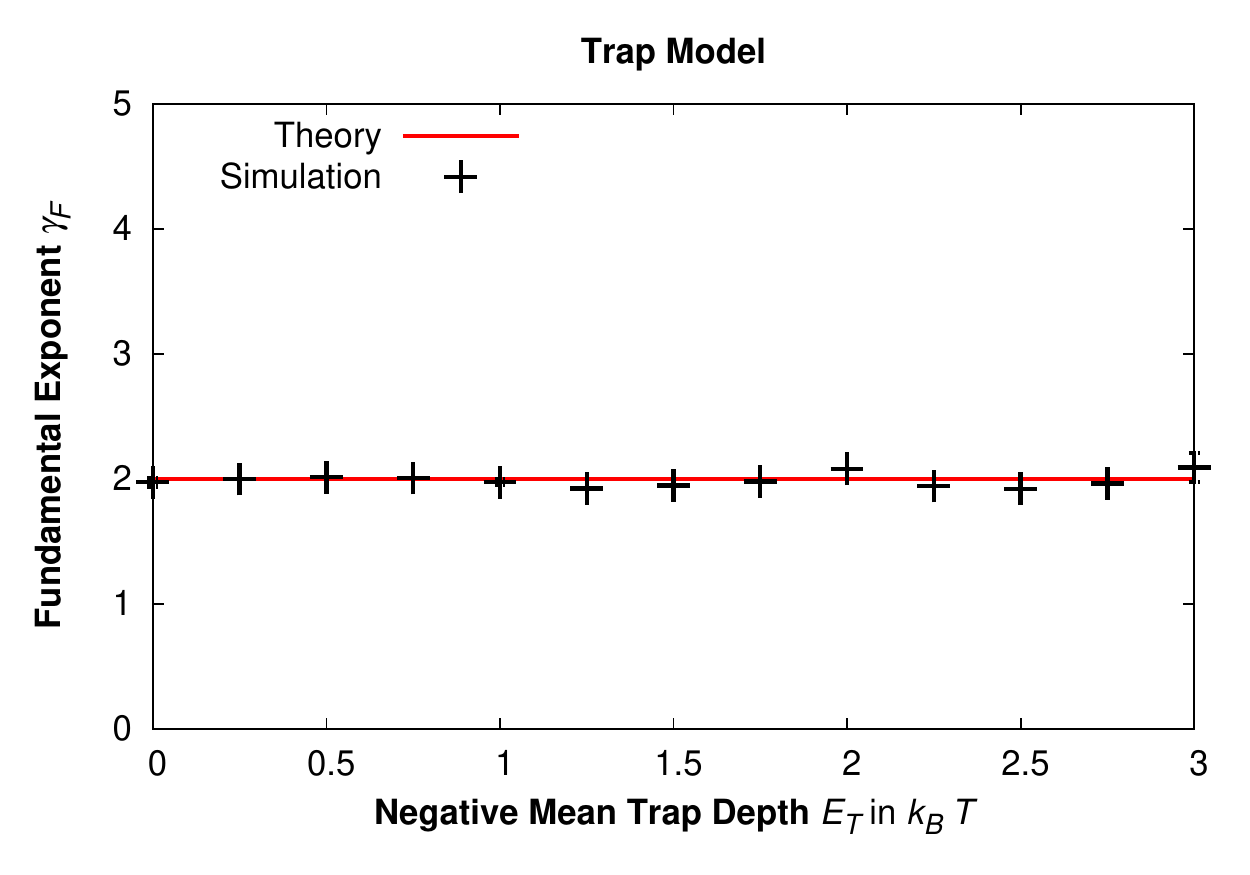}
			\begin{caption}{
				Simulation results for CTRW with Pareto distributed waiting times (upper panel) and for the trap model (lower panel).
				Every point is calculated from $2048$ trajectories of length $10^6 \tau$.
				Error bars are given for three points and are smaller than the symbols.
				Both models are (quasi-) subordinated to a process with uncorrelated increments, and
				their fundamental exponent is $\gamma_F = 2$.}
				\label{CTRWTrapFig}
			\end{caption}
		\end{figure}

		An example of such a situation is given by the random trap model which in 3d is adequately modelled by CTRW 
			but in 1d is a process where the temporal component of the process is not independent on the positional one.
		Let us compare CTRW and the random trap model in 1d \cite{Burov2011}.
		In the trap model each site corresponds to a potential well with random depth $E_i < 0$.
		The distribution of $E_i$ is taken to be exponential and characterized by the mean trap depth $ - E_T$.
		The jumping rates to the neighboring sites $w_{i \rightarrow i \pm 1}$ depend on $E_i$ according to Kramers' law
		\begin{equation}
				w_{i \rightarrow i \pm 1} 
			=
				\frac{1}{2} w_0 \exp( E_i / k_B T ),
			\label{eq:TrapRate}
		\end{equation}
		and the probability to go left or right upon a jump are the same. 
		The value of $\tau = w_0^{-1}$ defines the time scale of the process and is sent to unity in what follows. 
		The mean sojourn time $\tau_i = 1/(w_{i \rightarrow i -1}+ w_{i \rightarrow i +1})$ differs from site to site. 
		In contrast to the trap model, the waiting times in CTRW are re-drawn after every jump 
			-- usually from some heavy-tailed (Pareto) distribution $p(\tau) \sim \tau^{-1-\alpha}$, with $\alpha \in (0,1)$ --  
			and not bound to a specific site.
		In both cases, however, the process is (quasi)-subordinated to a process with uncorrelated increments, 
			and the fundamental exponent is expected to be two.
		This finding is supported by numerical simulations, see Fig.~\ref{CTRWTrapFig}.

	\textbf{iii) Further applications to disordered systems.} 
		Let us now concentrate on a single particle moving in a generic disordered energy landscape. 
		On the coarse-grained level the particle's motion on a lattice is described by a master equation 
			with the transition rates $w_{ij}$ between the neighboring sites fulfilling the detailed balance condition in equilibrium. 
		Under quite general assumptions (see \cite{Camboni2012}) the effective diffusion coefficient in such a model is 
		\begin{equation}
			D^* = a^2  \frac{ \langle w_{ij} \exp ( - E_i / k_B T) \rangle_{EM} }{ \langle \exp ( - E_i / k_B T) \rangle }
		\end{equation} 
			with $a$ being the lattice constant, $E_i$ again describing the particle's energy at site $i$, 
			and $\langle \sigma_{ij} \rangle_{EM}$ denoting the procedure of averaging, giving the macroscopic conductance 
			of a lattice with condictivities $\sigma_{ij}$ of single bonds.
		Subdiffusion is observed, provided this coefficient of normal diffusion vanishes.
		It can vanish either when the numerator vanishes (i.e. when the system is on percolation threshold, or in 1d) 
			or if the denominator diverges. 
		Since the denominator is proportional to the mean waiting time on a site, 
			this corresponds to diverging waiting times, 
			i.e. the situation which may take place in trap models.

		The situation with diverging waiting time (i.e. diverging Boltzmann factor) is termed as energetic disorder.
		The situation when the enumerator vanishes, depending on the structure of the system, 
			will be called structural disorder. 
		This is typical for all percolation cases and for barrier models in one dimension.  

		The stochastic processes generated by these models in single realizations are rather difficult to treat.
		However, as we proceed to show via numerical simulations, as soon as an average over the disorder is applied, 
			the corresponding processes \textit{on the average}  behave pretty much like the processes 
			subordinated to processes with stationary incremets.
		The correct averaging procedure here corresponds to averaging over \textit{many} realizations of our 
			disordered system with \textit{exactly one} realization of a random walk process in each of them. 

\begin{figure}
			\begin{center}
				\includegraphics[width=0.4\textwidth]{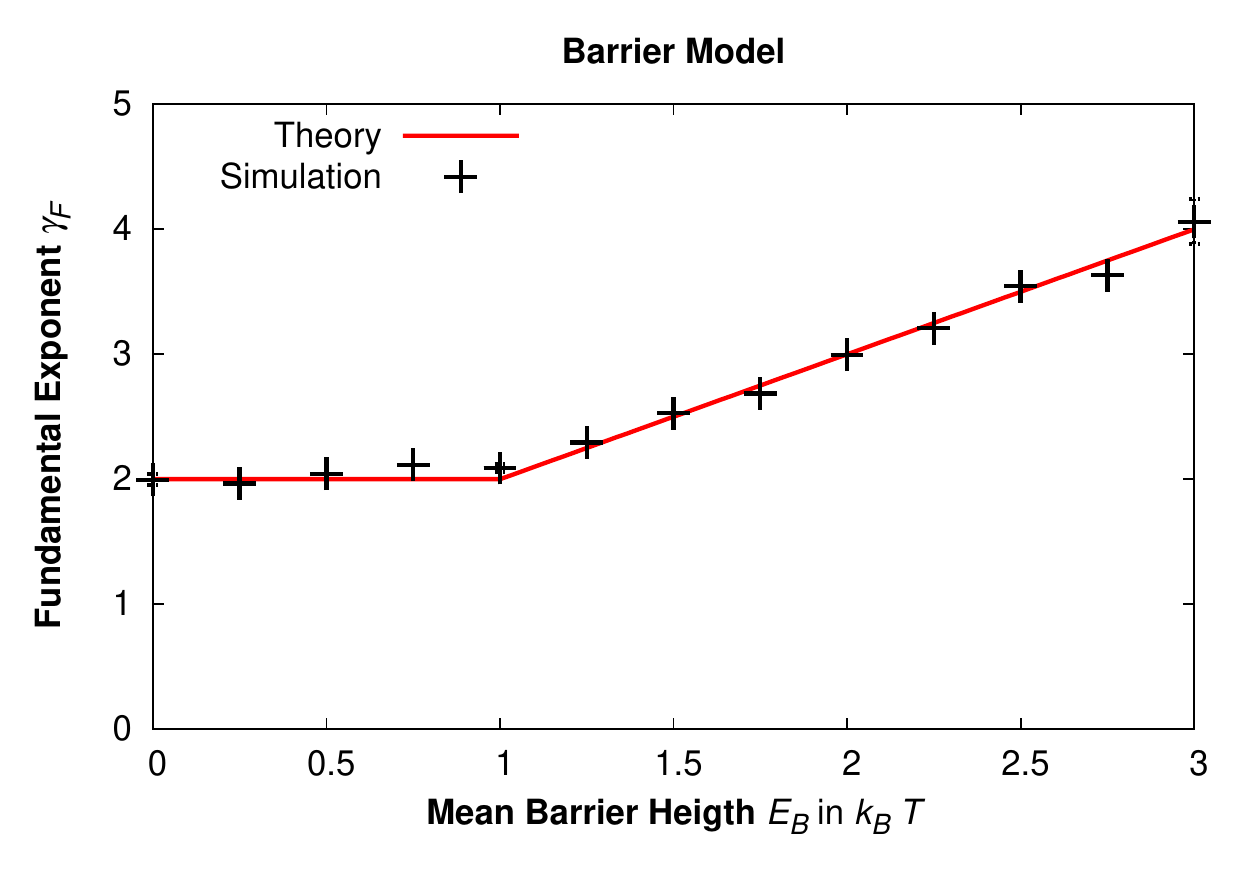}
				\includegraphics[width=0.4\textwidth]{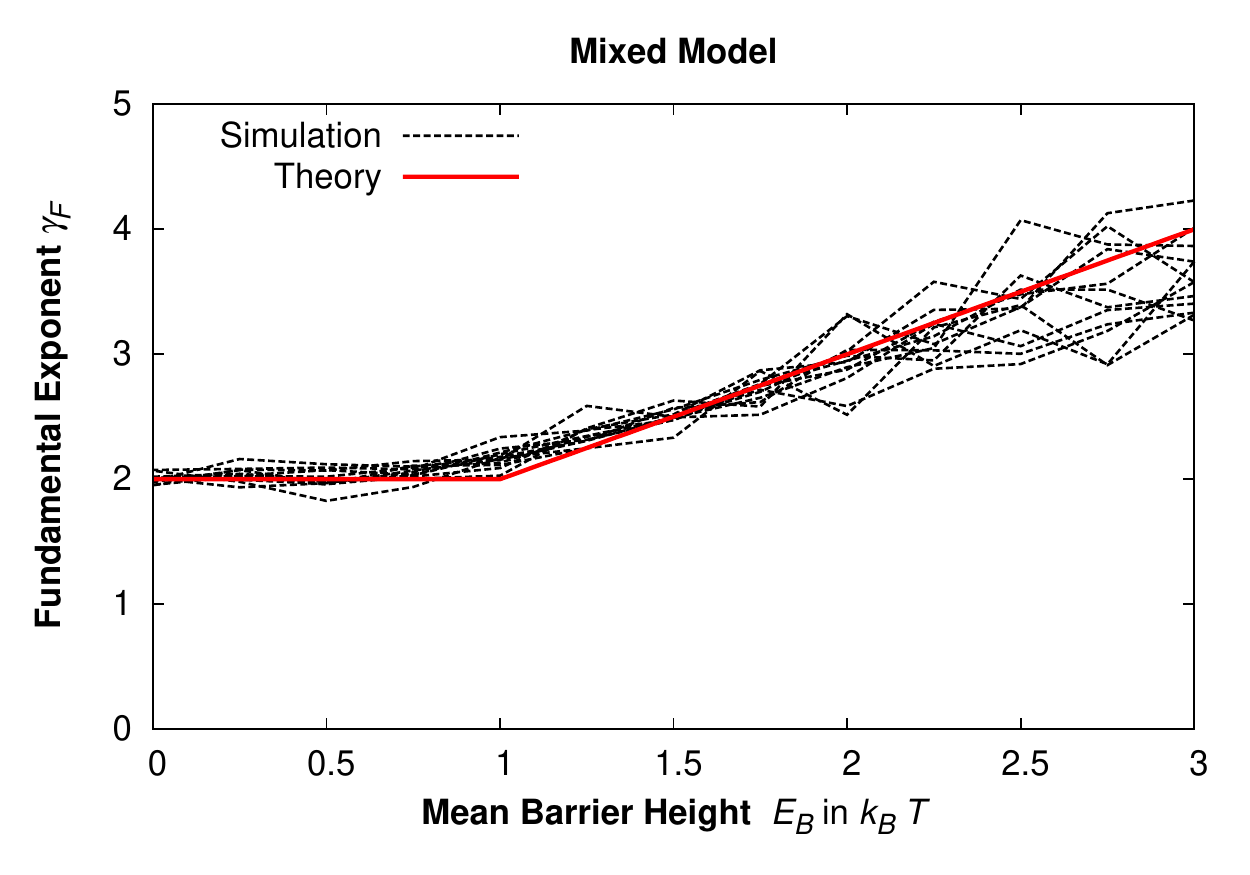}
			\end{center}
			\begin{caption}{Simulation of barrier (upper panel) and mixed (lower panel) models.
				The fundamental exponent is plotted against the mean barrier heigth. 
				Every point was calculated from $2048$ trajectories of length $10^{6}\tau$.
				Error bars are given for the barrier model for three points and have at most the size of the symbols.
				Each dotted line in the lower plot corresponds to one fixed value of mean trap depth varying from zero to $3 k_B T$.}
				\label{BarrMixdFig}
			\end{caption}
		\end{figure}

		As an example, let us examine the barrier model in 1d \cite{Haus1987,Bouchaud1990}. 
		This model is characterized by random energy barriers $E_{i,i+1} > 0$ placed between the sites of a chain.
		The heights of the barriers are exponentially distributed with $E_B$ being the mean height.
		The transition rates $w_{ij}$ are:
		\begin{equation}
			w_{i,i+1} = w_0 \exp( - E_{i,i+1} / k_B T ) 
			\text{.}
			\label{eq:BarrTrans}
		\end{equation}
		The random walks in each realization might be very different and may lack ergodicity, 
			but after averaging over all realizations 
			the process can be approximated by a possess with stationary increments:
			upon averaging all sites become equivalent, 
			and therefore the further displacement cannot depend on where (and therefore on \textit{when}) the process started. 
		Thus, the fundamental moment in our our barrier system has the index $\gamma_F = 2/\alpha$ with $\alpha$ being the exponent of the anomalous diffusion
		\[
			 	\alpha 
			= 
				\left\{
					 \begin{array}{ll}
						 \frac{2 k_B T}{k_B T + E_B} & \mbox{for   } E_B > k_B T \\
						 1 & \mbox {otherwise},
					 \end{array}
		 		\right.
		\]
		 (see \cite{Bouchaud1990}) as if the process were a pure (non-subordinated) process with stationary increments.
		The statements above are confirmed by results of numerical simulations in Fig.~\ref{BarrMixdFig}

		Now let us turn to a generic one-dimensional random potential being a combination of barriers and traps. 		
		The transition rates in such a process read
		\begin{equation}
			w_{i,i+1} = w_0 \exp( - (E_{i,i+1} - E_i) / k_B T ).
			\label{eq:MixdTrans}
		\end{equation}
		Note that in the cases pertinent to normal diffusion, when Eq.(4) holds, the numerator would depend only on $E_{i,i+1}$
		(i.e. on the properties of the barriers) and denominator only on $E_i$ (the properties of the traps). 
		We proceed to show (by means of numerical simulations) that the corresponding distinction is still possible in the subdiffusive domain.
		
		Averaging over the realizations of disorder leads us to a parent process with stationary increments 
			(by implicit averaging over barriers as discussed above), 
			and at the same time destroys correlations between the waiting times on the sites. 
		The whole situation can thus be approximated by a subordination of a parent process pertinent to averaged barrier behaviour 
			to a leading process stemming from the corresponding trap model. 
		This is a true process of anomalous diffusion of mixed origins. 
		Since the quasi-subordination introduced via the energy traps does not alter the value of $\gamma_F$, 
		we expect it to be $\gamma_F = 1 + E_B/(k_B T)$.
		The result of our numerical simulations are shown in the lower panel of Fig.~\ref{BarrMixdFig}. 
		In this figure we compared the numerical results of the barrier models with those of the mixed situation,
			where no systematic influence of the mean trap depth on the fundamental exponent is found.
		The only difference between the models are stronger fluctuations of $\gamma_F$ in the mixed model.
		
\begin{figure}
			\includegraphics[width=0.4\textwidth]{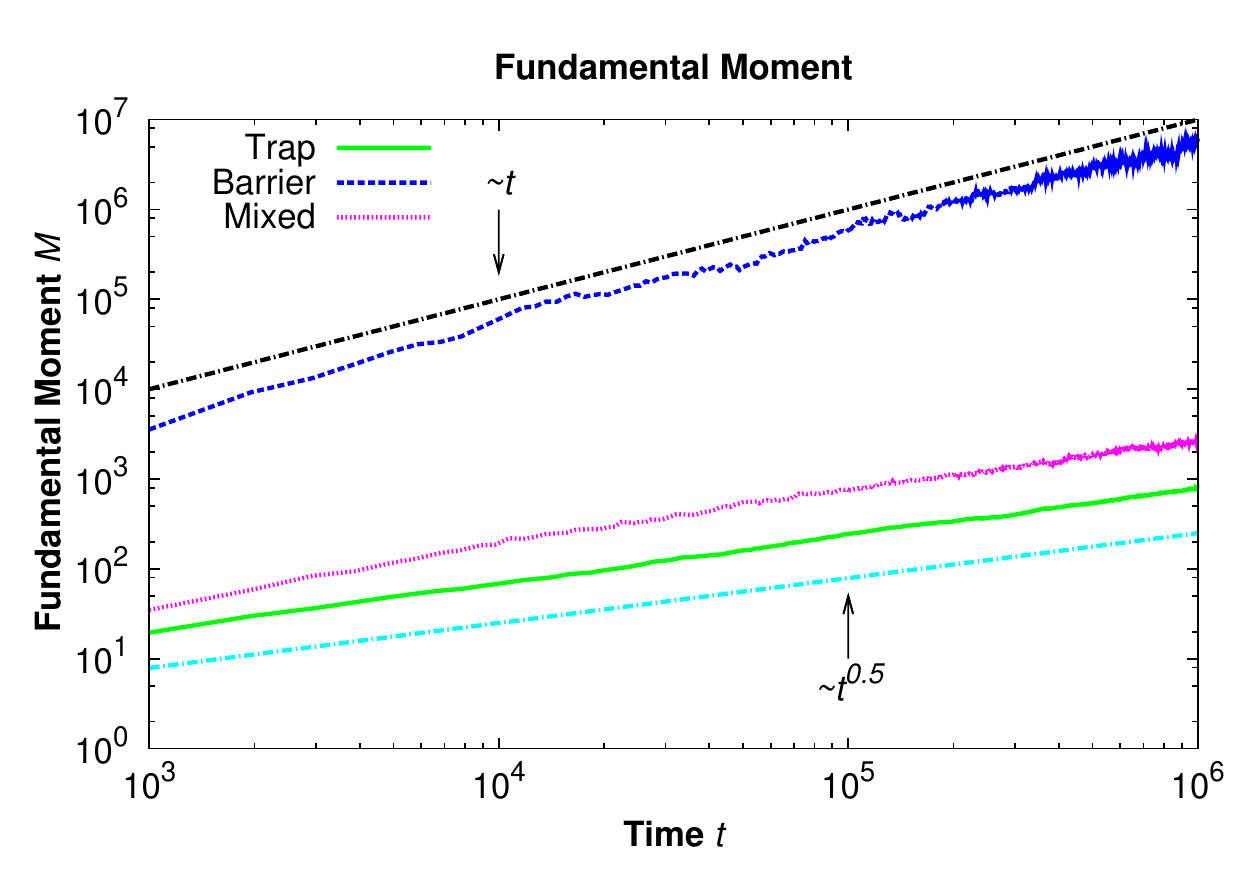}
			\begin{caption}{
				The fundamental moments for the trap, barrier and mixed models as functions of time.
				Here we took $E_T=3$ and $E_B=3$, so that both in the barrier and in the trap model $\langle X^2(t) \rangle \propto \sqrt{t}$.
				The upper and the lower dashed lines have slopes 1 and 0.5, respectively. }  
				\label{FundMom}
			\end{caption}
		\end{figure}

		In Fig.~\ref{FundMom} we plotted the fundamental moments of the barrier, trap and mixed model as functions of time.
		One observes that the slope of the mixed model follows the one of the trap model with the same distribution of potential wells' depth.
		This supports our statement made above: 
		The fundamental exponent is governed by structural disorder, while the time-dependence of the fundamental moment is governed by the energetic one. 
		
		To summarize: subdiffusion may stem from different physical mechanisms, 
			and distinguishing between them (or identifying combinations thereof) is an important task. 
		Such distinctions can be made on the basis of fundamental moment of the process, i.e. its absolute moment which is additive in time. 
		In subordinated processes, the index of the fundamental moment is inherited from the parent process and its time-dependence from the leading one. 
		In models of disordered potentials, the index is governed by the structural part of the disorder while the time dependence is given by its energetic part. 
		\acknowledgements{The authors acknowledge financial support by DFG within IRTG 1740 research and training group project. 

	\bibliographystyle{aipnum4-1}
	\bibliography{fund}
\end{document}